\documentclass[preprint,prb,showpacs,amsmath,amssymb]{revtex4}
\usepackage{graphicx}
\usepackage{dcolumn}
\usepackage{bm}
\usepackage{amsfonts}
\usepackage{amsmath}
\usepackage{amssymb}

\begin{document}
\title{Charge carrier injection into insulating media: single-particle versus 
mean-field approach}
\author{Yu.A.~Genenko}
\email{yugenen@tgm.tu-darmstadt.de}
\author{S.V.~Yampolskii} 
\author{C.~Melzer} 
\author{K.~Stegmaier}
\author{H.~von Seggern}   %
\affiliation{Institute of Materials Science, Darmstadt University of Technology, 
Petersenstra{\ss}e 23, 64287 Darmstadt, Germany}%
\date{\today}

\begin{abstract}

Self-consistent, mean-field description of charge injection into a dielectric 
medium is modified to account for discreteness of charge carriers. The 
improved scheme includes both the Schottky barrier lowering due to the individual
image charge and the barrier change due to the field penetration into the injecting
electrode that ensures validity of the model at both high and low injection rates
including the barrier dominated and the space-charge dominated regimes. Comparison
of the theory with experiment on a unipolar indium tin oxide/poly(phenylene vinylene)/Au device is presented. 

\end{abstract}

\maketitle

\section{\label{sec:intro}Introduction}

Injection of charge carriers from a conductor to an insulating medium has a crucial 
impact on the functioning of a wide range of electronic devices: light-emitting diodes, 
field-effect transistors, capacitors etc. using inorganic or organic semiconductors, 
wide-gap insulators and ferroelectrics. Nevertheless, the description of this process 
still remains controversial. By the simulation of inorganic and organic semiconductor 
devices the fixed values of charge carrier densities at the boundary with the insulating 
region are often used as boundary 
conditions~\cite{Emtage1966,Hughes1981,Hack1985,Shen1998,Cech2000,Mensfoort2008}. For 
charge transport in bulk dielectrics the electric field  at the injecting 
interface is often taken equal to zero assuming space-charge limitation of the 
current~\cite{Lambert,Christen2007,Fitting2007,Neumann2005}. Given finite values of 
the electric field at the interface are supposed in numerous papers simulating injection 
as Fowler-Nordheim (FN) tunneling through the surface energetic 
barrier~\cite{Blom1996,Malliaras1998,Walker2002,Martin2002,Martin2005} 
or Richardson-Schottky (RS) thermionic emission over this 
barrier~\cite{Walker2002,Davids1997,Campbell1998,Arkhipov1998,Arkhipov1999,%
Malliaras1999,Burin2000,Tessler2000,Ruhstaller2001,Preezant2003,Hosseini2005}. However, 
the problem is that the field and the carrier density at the interface are, in fact, 
unknown.

Moreover, the injection treatment in the spirit of the FN or the RS approach applies, 
strictly speaking, only to the case of a  single particle penetration since the 
presence of other particles immediately changes the electrostatic potential profile 
used in these models. This makes the application of the mentioned models questionable,
particularly at high injection levels. 
A consequence of the single-particle approach is that the characteristics 
of a dielectric medium are virtually absent in the results. Expressions for the FN 
and RS injection currents look as if it is injected into vacuum (except for the 
dielectric permittivity) and do not depend either on density of states available 
or on the charge carrier mobility in the dielectric medium, that contradicts 
experiment~\cite{Shen2001}. This deficiency of the RS model was properly noticed in 
Refs.~\cite{Emtage1966,Crowell1966} where the drift-diffusion equation for charge
carriers in the insulating medium was combined with RS injection to introduce the 
carrier mobility and the density of states in the injection current, however, the 
question remains whether a boundary condition derived from a single-particle model 
may be coupled with many-particle equations. In view of disordered insulators, 
sophisticated extensions of the RS approach describing injection as random hopping of 
a charge carrier in the same potential profile with account of possible injection
into the tail states below the barrier were developed~\cite{Arkhipov1998,Arkhipov1999,%
Burin2000,Arkhipov2003,Reynaert2004,Woudenbergh2005} which, however, do not fix the 
general deficiency of the model of injection which remains a single-particle one.

For a proper description of the charge carrier injection a self-consistent determination 
of the field and the carrier density at the conductor/insulator interface is necessary 
which would contain both limits of weak (single-particle) and strong (many-particle) 
injection including the space-charge limit. Recently, a self-consistent continuous
description of injection in terms of carrier densities and mean fields was 
advanced~\cite{Neumann2006,Neumann2007} based on the matching of the 
electric displacement and the electrochemical potential at the interface. 
This one-dimensional treatment implies averaging of all variables over the plane 
perpendicular to the injection current density which pertains to relatively high 
carrier densities. The model exhibits a plausible crossover from the barrier dominated 
behavior at low voltages to the space-charge dominated behavior at high voltages and 
also a field-induced reduction of the injection barrier. Nevertheless, at medium 
voltages, this reduction is much less than the reduction of the Schottky barrier due 
to the single-particle image potential of three-dimensional nature~\cite{Sze} which 
seems to have been confirmed in experiments~\cite{Walker2002,Davids1997,Campbell1998}. 
Apparently, there are restrictions on application of the mean-field description of 
the charge injection at low carrier densities.

The importance of proper account of the discrete, single-particle nature of the 
interaction between an injected carrier and the injecting electrode was emphasized in 
the Refs.~\cite{Bussac1998,Tutis1999,Tutis2001} where a sophisticated numerical 
treatment of the one-dimensional hopping transport in organic semiconductors was developed 
including the contribution of the modified single-particle image charge potential. 
However, for the boundary conditions in this model, the tunneling from an electrode was 
employed assuming numerous atomistic parameters of the involved media instead of the 
density of states available and the carrier mobility.

The most advanced approach to charge injection including the Schottky barrier lowering 
and the space charge effect was recently developed in Ref.~\cite{Holst2009} where 
three-dimensional (3D) hopping of charge carriers over the sites of a cubic lattice with 
randomly distributed energy levels was considered as well as a sophisticated 
one-dimensional (1D) continuous model. The site occupancies and the electric 
field were calculated self-consistently by solving the three-dimensional master 
equation and the Poisson equation in consequent iterations with account of the 
field-dependent injection barriers. On the other hand, the contribution of the individual 
image potential into injection was accounted inconsistently which lead to an 
overestimation of the image effect as will be shown below.

The primary aims of this paper are to establish the requirements, which allow for the 
mean-field treatment of charge injection, and to extend the boundary conditions for 
the continuous description~\cite{Neumann2006,Neumann2007,Shashkin2008} in such a way that 
the effect of discreteness of charge carriers is considered. We first look for a simple 
criterion separating the regions of system parameters (voltage and injection barrier) 
where either the single-particle 
(SP) or the many-particle, mean-field (MF) concepts are valid. Then we formulate 
self-consistent boundary conditions which account for the discrete character of the 
particle interaction with the electrode and provide the crossover from low to high 
injection regime including the case of space-charge domination. Finally, the results are 
compared with an experiment.

\section{\label{sec:generalmodel}Comparing a single-particle and a mean-field concepts}

Adequacy of the MF description of the charge carrier injection or, alternatively, of the 
SP picture is determined mostly by the relation between two respective characteristic 
lengths. In the first approach this length is given by an average, three-dimensional 
distance between the injected charge carriers defined as $r_{s}=p_s^{-1/3}$ where $p_s$ 
is the density of these carriers. It is assumed here that the distribution of the 
particles is homogeneous at least over the distance of few $r_{s}$ in all space 
directions. In the second approach the relevant length is the distance $x_{m}$ from the plane 
surface of an electrode to the position of the potential maximum of the Schottky 
energetic barrier~\cite{Sze}.

We verify now the validity of the SP approach from the point of view of the MF 
approximation and vice versa. If the density of injected carriers $p_s$ predicted in the 
continuous consideration is so high that $r_{s}\ll x_{m}$ then the presence of many 
particles between the surface of the electrode and the maximum of the single-particle 
potential makes the SP approach inappropriate; {\it nota bene} that MF model density 
$p_s$ is used in this estimation because in the consistent SP calculations the density 
of injected carriers makes no sense. If, on the other hand, $x_{m}\ll r_{s}$ then the 
interaction of a single injected charge carrier with its image is much stronger than 
with the other distant injected carriers which makes the MF approximation inappropriate. 
At first sight, the criterion $x_{m}\simeq r_{s}$ must roughly  separate the regions of 
relevance of the discrete SP and the continuous MF concepts~\cite{Genenko2008}. In fact, 
comparison of the SP image force exerted upon an injected carrier with random forces due 
to other charge carriers needs more sophisticated analysis and involves additionally 
other characteristic lengths. To perform this analysis we first delineate below the SP 
and the MF approaches and then study their marginal validity. To make this delineation 
quantitative we consider in the following exemplarily injection of holes from an 
indium tin oxide (ITO) electrode into an organic semiconductor.

\subsection{\label{subsec:density}Boundary conditions in the mean-field approach}

Here the treatment in terms of continuous carrier densities and fields is assumed.
In the linearized Thomas-Fermi approximation~\cite{Ashcroft} the electric field
in the electrode, occupying the space at $x<0$, reads~\cite{Neumann2006}
\begin{equation}
\label{conductor-field}
F_{c}(x)=\left[ F_c(0) - \frac{j}{\sigma_c} \right]
\exp{\left(\frac{x}{l_{TF}}\right)} + \frac{j}{\sigma_c},\,\,\,x<0, 
\end{equation}

\noindent where $j$ is the position-independent steady-state current density, 
$\sigma_c$ the conductivity of the electrode, and 
\begin{equation}
\label{LTF}
l_{TF}=\sqrt{\frac{2\epsilon_0\epsilon_c \kappa_{\infty}}{3e^2 p_{\infty }}}
\end{equation}
\noindent is the Thomas-Fermi screening length. Here $e$ is the positive elementary 
charge, $\epsilon_0$ the dielectric permittivity of vacuum, $\epsilon_c$ the static 
relative permittivity of the electrode, $p_{\infty }$ and $\kappa_{\infty}$ are, 
respectively, the equilibrium values of the carrier density and of the chemical 
potential in the electrode far away from the interface~\cite{Dawber}.

The electrochemical potential in the electrode equals~\cite{Neumann2006}
\begin{equation}
\label{chem-pot-electrode}
\kappa_{c}(x)=E_b+\kappa_{\infty} + e l_{TF} \left[ F_c(x)-\frac{j}{\sigma _c}\right] 
+ e\phi_c (x),\,\,\,x<0, 
\end{equation}
\noindent where $E_b$ terms the position of the conduction band bottom and
$\phi_c (x)$ the electrostatic potential in the electrode.

The electric field in the organic semiconductor, $F_s(x)$, obeys 
equation~\cite{Walker2002,Sze}
\begin{equation}
\label{organic-eq}
\frac{kT}{e} F_s''(x)-F_s(x)F_s'(x)=-\frac{j}{\mu_s\epsilon_0\epsilon_s},\,\,\,x>0,
\end{equation}
\noindent where $\mu_s$ denotes the hole mobility in the organic semiconductor, 
$\epsilon_s$ its static relative permittivity, $k$ the Boltzmann constant and $T$ the 
absolute temperature. Nonlinear equation (\ref{organic-eq}) is usually solved 
numerically~\cite{Walker2002} but can also be solved 
analytically~\cite{Neumann2007,Shashkin2008} which offers a certain advantage as can be 
seen below. The solution reads
\begin{align}
\label{solspecial}
&F_s(x) =-\frac{2kT}{e l_{TF}}\iota^{1/3} \\
&\times
\frac{ \mathrm{Ai}^{\prime}\left[\iota^{1/3}
\left(\frac{\displaystyle x}{\displaystyle l_{TF}}+C_1\right)\right] +
C_2\mathrm{Bi}^{\prime}\left[\iota^{1/3}
\left(\frac{\displaystyle x}{\displaystyle l_{TF}}+C_1\right)\right] }
{ \mathrm{Ai}\left[\iota^{1/3}
\left(\frac{\displaystyle x}{\displaystyle l_{TF}}+C_1\right)\right] +
C_2\mathrm{Bi}\left[\iota^{1/3}
\left(\frac{\displaystyle x}{\displaystyle l_{TF}}+C_1\right)\right] },
\nonumber
\end{align}
\noindent where $\mathrm{Ai}$ and $\mathrm{Bi}$ denote the Airy 
functions~\cite{Abramowitz}, and 
$\iota=je^2l_{TF}^3/(2\mu_s\epsilon_0\epsilon_s (kT)^2)$. For the semi-infinite 
geometry considered in this section the constant $C_2$ should be chosen equal to zero 
since the charge carrier density vanishes asymptotically~\cite{Neumann2007} whilst the 
constant $C_1$ must be determined from the boundary condition at the interface.

Assuming Boltzmann statistics for charge carriers in the organic semiconductor we 
proceed, similar to Sze for inorganic semiconductors~\cite{Sze}, with the 
electrochemical potential 
\begin{equation}
\label{chem-pot-organic}
\kappa_{s}(x)=E_b+\kappa_{\infty} + \Delta + kT \ln{\left[ \frac{p_s(x)}{N}\right]}    
+ e\phi_s (x),\,\,\,x>0, 
\end{equation}
\noindent where the zero-field barrier $\Delta$ is here given by the difference between 
$E_b+\kappa_{\infty}$ and the highest occupied molecular orbital (HOMO) in the organic 
medium, $N$ is the density of states for holes at the HOMO-level, $p_s(x)$ the density 
of holes and $\phi_s (x)$ the electrostatic potential in the organic medium. In fact,
in disordered (organic) semiconductors the injection barrier becomes an ill-defined
quantity because such systems do not exhibit sharp band edges. The barrier is then often
defined as a difference between the Fermi level in the injecting electrode and the 
center of the density of states (DOS) distribution  at the HOMO-level which is typically
of the Gauss type~\cite{Mensfoort2008,Arkhipov1998,Arkhipov1999,Burin2000,Arkhipov2003,%
Woudenbergh2005,Holst2009}. Using in the following the narrow-band approximation we 
assume a negligible width of the mentioned Gauss DOS $\sigma \ll \Delta$. We note, on 
the other hand, that the following results apply for wide-band non-degenerate 
semiconductors and insulators as well in which case $2(2\pi mkT/h^2)^{3/2}$ 
with $m$ the effective mass of carriers, and $h$ the Planck constant should be taken for 
the effective density of states~\cite{Sze,Shashkin2008}.

To obtain the boundary condition at the interface at $x=0$ we assume that 
\begin{enumerate}
\item 
electrostatic potential is continuous, $\phi_c (-0)=\phi_s (+0)$, which means absence of 
a dipole layer at the interface; 
\item 
electric displacement is continuous, 
$\epsilon_c F_c(-0)=\epsilon_s F_s(+0)$, which implies absence of a surface 
charge at the interface;
\item 
electrochemical potential is continuous through the system, which provides
matching at the interface, $\kappa_{c}(-0)=\kappa_{s}(+0)$. 
\end{enumerate}
\noindent Note that the requirements 1. and  2. account for the image charges of all 
injected carriers in the mean-field approximation. The conditions 1.-3. together 
deliver the boundary condition at the interface~\cite{Neumann2006,Neumann2007}
\begin{equation}
\label{boundary-cond-single}
\frac{p_{s}(0)}{N} = \exp{\left[ -\frac{\Delta }{kT} + \frac{el_{TF}}{kT}
\left(  \frac{\epsilon_s }{\epsilon_c} F_s(0)  - \frac{j }{\sigma_c} \right) \right]},
\end{equation}
\noindent which connects the values of the field and the carrier density in the organic
semiconductor at the interface. Due to presence of the current this boundary condition 
describes, in principle, a non-equilibrium state of the contact but in fact the 
contribution of the current in Eq.~(\ref{boundary-cond-single}) can be neglected here 
because it is very small in all practical cases concerning organic 
semiconductors~\cite{Neumann2006,Neumann2007}. That is why it will be omitted in the 
following modifications of the formula~(\ref{boundary-cond-single}). We note also that 
the condition~(\ref{boundary-cond-single}) does not define explicitly the values of the 
field and the carrier density at the interface; these quantities can only be found by 
satisfying Eq.~(\ref{boundary-cond-single}) together with Eq.~(\ref{solspecial}).

Since Boltzmann statistics is assumed to calculate the chemical potential in the organic 
medium the argument of the exponential function in the 
formula~(\ref{boundary-cond-single}) must be negative which imposes a restriction on the 
field value $F_s(0)<F_{lim}=\epsilon_c\Delta/\left(\epsilon_s el_{TF}\right)$. 
The form of Eq.~(\ref{boundary-cond-single}) implies an effective barrier for hole 
injection 
\begin{equation}
\label{barrier-eff}
\Delta_{eff}=\Delta - eF_s(0)l_{TF}\epsilon_s/\epsilon_c = \Delta - eF_c(0)l_{TF},
\end{equation}            
\noindent which states a barrier modification due to the electric field at the 
interface. The barrier correction in Eq.~(\ref{barrier-eff}) is linear in field and may 
be positive or negative depending on the sign of the local field at the interface. The 
physical reason for this barrier modification  is a contribution of the work done by the 
mean electric field on electrons in the electrode.

\subsection{\label{subsec:Schottky}Schottky barrier and image potential in the 
single-particle approach}

In the classical description of a Schottky barrier~\cite{Walker2002,Arkhipov2003,Sze}, 
the barrier for the emission of a single particle through a plane interface is formed by 
the superposition of the zero-field barrier $\Delta$, the potential of attraction to the 
image charge of the opposite sign at the electrode and the contribution of the mean 
electric field in the organic semiconductor $F_s(x)$: 
\begin{equation}
\label{single-barrier}
U(x)=\Delta -\frac{e^2}{16\pi\epsilon_0\epsilon_s x} - e F_s(0)x,\,\,\,x>0, 
\end{equation}
\noindent where $x$ denotes the distance from the ITO/organic interface, and $F_s(0)$ 
the value of the electric field at the interface. The electric field $F_s(x)$ 
is assumed positive and virtually constant within the distance $\approx x_{m}$ from the 
interface, since for the negative field the potential exhibits no maximum and does not 
allow particle escape from the potential well. The maximum of the potential barrier,
Eq.~(\ref{single-barrier}), is located at a position 
\begin{equation}
\label{x3D}
x_{m}=\sqrt{\frac{e}{16\pi\epsilon_0\epsilon_s F_s(0) }}
\end{equation}
\noindent and amounts to $U(x_{m})= \Delta -e\delta \phi_{Sch}$ with
\begin{equation}
\label{Schottky-lowering}
\delta \phi_{Sch}=  2 F_s(0)x_{m} =           
\sqrt{\frac{e F_s(0)}{4\pi\epsilon_0\epsilon_s }},
\end{equation}
\noindent where the latter provides the so-called Schottky-barrier lowering~\cite{Sze}. 
This barrier modification is proportional to the square root of the field and always 
reduces the barrier in contrast to the barrier correction in Eq.~(\ref{barrier-eff}) 
which is linear in the field and may have different signs. The physical reason for the 
Schottky-barrier lowering is also different: in contrast to the contribution of the mean 
electric field at the side of the electrode as described by Eq.~(\ref{barrier-eff}), 
Eq.~(\ref{Schottky-lowering}) results from the energy profile modification in the 
dielectric medium due to the individual image-charge effect.

Eq.~(\ref{single-barrier}) implies ideal screening of the electrostatic field by the
metal surface at $x=0$. ITO is, however, a highly doped semiconductor which does not
provide ideal screening of the field. Taking into account the dielectric permittivity 
of the electrode $\epsilon_c$ and the characteristic length of the field penetration 
into the electrode $l_{TF}$ the energy of the injected charge carrier is substantially 
modified~\cite{Kornyshev1977}. While at large distances from the interface 
$x\gg a=(\epsilon_s/\epsilon_c)l_{TF}$ the energy approaches asymptotically  
Eq.~(\ref{single-barrier}), at small distances $x\ll a$, the Coulomb term  
$\sim e^2(\epsilon_s-\epsilon_c)/((\epsilon_s+\epsilon_c)16\pi\epsilon_0\epsilon_s x)$
prevails, which can even change sign from attraction to repulsion depending on the
magnitudes of respective permittivities. At such small distances a consistent 
quantum-mechanical treatment of injection becomes necessary which, in turn, essentially 
modifies the energy profile in the close vicinity of the 
interface~\cite{Gabovich1979,Voitenko1981}. For our consideration, the important maximum
of the potential~(\ref{single-barrier}) would fall into the region $x_m\sim a$ for a 
voltage of about $eL/16\pi\epsilon_0\epsilon_s a^2$ with $L$ the device thickness. 
For typical parameters of both involved media given in the 
Table~\ref{Materialparameters1} and $L\simeq$100 nm this voltage amounts to 200 V which 
is far too large for organic materials. Therefore in the following the classical 
approximation for the image potential, Eq.~(\ref{single-barrier}), will be used keeping 
in mind possible modifications of this interaction when other electrode and dielectric 
materials are involved.

\begin{table*}
\caption{\label{Materialparameters1}Typical material parameters for the injecting 
electrode and an organic semiconductor 
(Refs.~\onlinecite{Martin2005,Ashcroft,Mergel2002,Mergel2004,Fujiwara2005}). 
The parameters are deduced assuming $T=300$~K.}
\begin{ruledtabular}
\renewcommand{\arraystretch}{1.5}
\begin{tabular}{cccccc|cccc}
\multicolumn{6}{c|}{ITO} & \multicolumn{4}{c}{Organic semiconductor} \vspace{2pt}\\
\hline
$l_{TF}$ & $p_{\infty} $ & $\epsilon_c $ & $ \mu_c $ & $\kappa_{\infty} $ & $E_{A}$ &
${N} $ &  $\epsilon_s $ & $ \mu_s $  & $E_{\text{HOMO}}$ \vspace{2pt}\\
($\mathring{\text A}$) & (cm$^{-3}$) &  & $\left( \displaystyle \frac{\text{cm}^2}{\text{V s}}\right)$ & (eV) & (eV) &
(cm$^{-3}$) &     &  $\left( \displaystyle \frac{\text{cm}^2}{\text{V s}}\right)$
 & (eV)  \vspace{2pt}
\\
\hline
8.6 & 10$^{20}$ & 9.3 & 30 & 0.225 & 4.7 & 10$^{21}$ &  3  & 10$^{-4}$  &  5.0 
\end{tabular}
\end{ruledtabular}
\vspace{0.5cm}
\end{table*}

\subsection{\label{subsec:criteria}Microscopic comparison of the single-particle image 
force and stochastic forces}

To establish on what terms the SP approach transforms to the MF one we first generalize 
the formula (\ref{single-barrier}). In classical statistical mechanics the microscopic 
density of injected particles is defined as~\cite{Klimontovich1986}
\begin{equation}
\label{micro-dens}
p(\mathbf{r},t) = \sum\limits_{i} \delta (\mathbf{r}-\mathbf{r}_i(t)),
\end{equation}
\noindent where $\delta (\mathbf{r})$ is the three-dimensional Dirac delta-function and
vectors $\mathbf{r}_i(t)$ indicate random positions of all injected particles and their 
images. The energy of an injected particle is then given by
\begin{equation}
\label{micro-energy}
U_M(\mathbf{r},t) =  U_0(\mathbf{r}) - \frac{e^2}{16\pi\epsilon_0 \epsilon_s  x} 
+\int d\mathbf{r'} p(\mathbf{r'},t)
\frac{e^2 \,\text{sgn}(x')}{4\pi\epsilon_0 \epsilon_s |\mathbf{r-r'}|}, 
\end{equation}
\noindent
and the $x-$component of the microscopic force exerted upon this particle reads 
\begin{align}
\label{micro-force}
f_M^x(\mathbf{r},t) & = -\frac{\partial U_0}{\partial x} 
-\frac{e^2}{16\pi\epsilon_0 \epsilon_s  x^2} \nonumber\\
& -\frac{\partial }{\partial x}\int d\mathbf{r'} p(\mathbf{r'},t)
\frac{e^2\,\text{sgn}(x')}{4\pi\epsilon_0 \epsilon_s |\mathbf{r-r'}|}, 
\end{align}
\noindent
where the first term in both Eqs.~(\ref{micro-energy}) and (\ref{micro-force})
presents the contribution of an external field, the second term is due
to the individual image charge and the third one accounts for the other injected 
particles. The signum function $\text{sgn}(x')$ in the last term accounts for opposite 
sign of the image charges. To judge whether the SP approach is relevant, the impact of
the deterministic individual image effect should be compared with the impact of the 
other particles. Due to the stochastic nature of the interaction of the individual 
particles leading to the energy and force fluctuations the comparison must be 
performed in two ways: the mean value or, alternatively, the variance
of the third term in Eq.~(\ref{micro-energy}) or Eq.~(\ref{micro-force})
should be compared with the second term. As long as the mean value and the variance
of the stochastic term is much less than the individual image charge contribution the
SP approach prevails.

\subsubsection{Mean-force criterion}

By configurational averaging of Eq.~(\ref{micro-force}) over all possible charge carrier
positions the mean, macroscopic density of particles becomes a continuous function of 
the only variable $x$: $<p(\mathbf{r},t)> = p_s(x)$. After integration over the space 
variables $y'$ and $z'$ the mean force results in the form
\begin{align}
\label{macro-force}
<f_M^x(\mathbf{r},t)> &= -\frac{\partial U_0}{\partial x} 
-\frac{e^2}{16\pi\epsilon_0 \epsilon_s  x^2}\nonumber\\ 
&+\frac{e^2 }{2\epsilon_0 \epsilon_s }
\int_{-\infty}^{+\infty} dx' p_s(|x'|) \text{sgn}(x') \text{sgn}(x-x'), 
\end{align}
\noindent
where the second and the third terms should be compared at $x=x_m$. Keeping in mind the
hopping transport over the random atomic or molecular sites typical of disordered 
(organic) semiconductors, we assume here that the characteristic length in the space
dependence of $p_s(x)$ is much larger than the nearest-neighbor distance between the
sites occupied by the charge carriers, $r_0\simeq 0.1 {-} 1\mbox{ nm}$.

Eqs.~(\ref{micro-energy})-(\ref{macro-force}) make sense if the density 
of injected particles decreases at distances much smaller than the device thickness, $L$, 
otherwise the electrostatic potential has to be modified to account for the other 
electrode as it is done in the Appendix. Nevertheless, even overestimation of the 
integral in Eq.~(\ref{macro-force}) for the very low constant concentration 
$p_s(x)\sim r_s^{-3}$ at $-L<x<L$ with $r_s\gg L$ results, by comparison of the second
and the third terms, in assessment $x_m \simeq \sqrt{r_s^3/8\pi L} \gg L$. Since such 
$x_m$ cannot be realized, it means that the third term in Eq.~(\ref{macro-force}) remains 
always much less than the second one, so that the SP contribution dominates in the limit
$r_s\gg L$. This formal statement is in agreement with the obvious fact that, for 
$r_s >L$, the description in terms of the continuous charge density is not valid.

For higher concentrations with $r_s \ll L$ we may apply to the 
evaluation of the integral in Eq.~(\ref{macro-force}) the equilibrium solution of 
Eq.~(\ref{organic-eq}) which is known to prevail in a wide range of applied 
fields~\cite{Neumann2006,Neumann2007}
\begin{equation}
\label{macro-density}
p_s(x) = p_s(0)/(1+x/\lambda\sqrt{2})^2,
\end{equation}
\noindent
with the Debye length $\lambda=\sqrt{\epsilon_0 \epsilon_s kT/e^2p_s(0)}$. By comparing
the second and the third terms in Eq.~(\ref{macro-force}) the criterion results
\begin{equation}
\label{crit1}
x_m = \frac{l_T}{2} + \sqrt{ \left( \frac{l_T}{2} \right)^2 
+ r_s\sqrt{\frac{r_s l_T}{16 \pi  }} },
\end{equation}
\noindent
where one more characteristic length $l_T = e^2/32\pi\epsilon_0 \epsilon_s kT$ 
$= r_s^3/32\pi\lambda^2$ 
of the order of the Coulomb capture radius~\cite{Scott1999} appears. For typical 
parameters of organic semiconductors at room temperature~\cite{Neumann2006,Neumann2007} 
$l_T$ is about 2 nm. In the range of concentrations $L^{-3}\ll p_s(0)\ll l_T^{-3}$ it 
follows from Eq.~(\ref{crit1}) in a good approximation that
\begin{equation}
\label{MF-criterion}
x_m \simeq r_s \left(\frac{l_T}{16\pi r_s}\right)^{1/4} 
\simeq (0.2{-}0.4) r_s. 
\end{equation}
This means that, for $x_m$ smaller than that given by Eq.~(\ref{MF-criterion}),
the SP term in the force, Eq.~(\ref{macro-force}), dominates while in the opposite 
case the SP contribution may be considered as embedded in the MF term.

For higher concentrations $p_s(0)\gg l_T^{-3}$ the equilibrium density 
(\ref{macro-density}) does not apply anymore. By increasing external field the injected 
charge density is known to transform from the diffusion induced equilibrium distribution 
(\ref{macro-density}) to virtually constant distribution over the 
device~\cite{Neumann2006} which transforms by further field increasing to the 
distribution $p_s(x)\sim 1/\sqrt{x}$ typical of space charge limiting 
currents~\cite{Lambert}.  Assuming that at $r_s \lesssim l_T$ the concentration 
$p_s(x) \sim l_T^{-3}$ remains constant all over the device and $x_m\ll L$ one 
obtains, by equating the second and the third term in Eq.~(\ref{macro-force}), the 
criterion $x_m \simeq r_s \sqrt{r_s/16\pi L} \simeq 10^{-1} l_T$. According to
Eq.~(\ref{x3D}) this can only be realized for very high fields where the classical 
SP approximation will be violated because of quantum effects discussed in 
Sec.~\ref{subsec:Schottky}. Thus, in the region of realistic electric fields the SP 
contribution remains much lower than the MF one so that the MF approximation prevails
for $r_s \lesssim l_T$. 

For even higher densities, when $r_s$ becomes much less than $l_T$, the MF approximation
of Sec.~\ref{subsec:density} is expected to fail because of violation of the criterion 
of applicability of Boltzmann statistics, $F_s(0)<F_{lim}$. On the other hand, the 
classical SP approximation might also be violated in this region because of the mentioned
quantum effects. In any case, this all happens at rather high voltages as will be seen in 
the following.

\subsubsection{Energy fluctuation criterion}

Considering the variation of the microscopic energy~(\ref{micro-energy}) we will need a 
second central moment of the microscopic carrier density which can be written in the 
form~\cite{Klimontovich1986}
\begin{equation}
\label{density-corr}
<\delta p(\mathbf{r'},t)\delta p(\mathbf{r''},t)>=
p_s(x) \delta (\mathbf{r'-r''}) + p^2_s(x) g_2(\mathbf{r'-r''}),
\end{equation}
with $\delta p(\mathbf{r},t)=p(\mathbf{r},t)-p_s(x)$ and $g_2(\mathbf{r})$ the 
correlation function. In terms of hopping over the molecular sites, this form assumes
that $p_s(x)$ changes on the scale much larger than both the distance between the 
nearest-neighbor sites, $r_0$, and the characteristic length of the correlations
contained in the function $g_2(\mathbf{r})$. In Refs.~\cite{Holst2009,Pasveer2005} an
assumption of the short-range correlations of the charge carrier positions was adopted 
which only excludes the double occupation of sites. This implies that 
$g_2(\mathbf{r})=-1$ when $|\mathbf{r}|<r_0$ and $g_2(\mathbf{r})=0$ otherwise as in the 
model of hard spheres~\cite{Klimontovich1986}. Such space correlations are compatible
with the hypothesis of the uncorrelated Gauss disorder of the energy levels at different 
sites in a random hopping system considered in 
Refs.~\cite{Burin2000,Holst2009,Pasveer2005}. The validity of the short-range space 
correlations of the particle positions can be verified by comparison with the screened 
Coulomb correlations given by the function~\cite{Klimontovich1986} 
\begin{equation}
\label{corr-Col}
g_2(\mathbf{r})=\exp{\left[ -\frac{8 l_T}{r}\exp{\left(-\frac{r}{\lambda } \right)} 
\right]} -1,
\end{equation}
which is also appropriate for the charged plasma of injected particles of the same sign. 
This formula defines indeed short-range correlations for medium and high charge
densities when $r_s\leq 10\, \mbox{nm}$ since in this case $\lambda \leq 2\, \mbox{nm}$.  
However, for lower particle concentrations with $r_s\simeq L/2 \simeq 50\, \mbox{nm}$
the correlation length becomes about $\lambda\simeq 20\, \mbox{nm}$. Other possible
correlations in organic systems and their effect on the results presented here will be 
discussed later in the concluding section.

The variation of the energy (\ref{micro-energy}) is due to the stochastic term only and can be estimated as
\begin{align}
\label{energy-variation}
&<\left( U_M(\mathbf{r},t)-<U_M(\mathbf{r},t)>\right)^2> & \nonumber\\
&\simeq \left( \frac{e^2}{4\pi\epsilon_0 \epsilon_s}\right)^2  
\int d\mathbf{r'} \frac{p_s(|x'|)}{(\mathbf{r-r'})^2}
\left[ 1 + p_s(|x'|)v_0  \right] \nonumber\\
&\simeq \left( \frac{e^2}{4\pi\epsilon_0 \epsilon_s}\right)^2 2\pi
\int_{-L}^{L} dx' p_s(|x'|) \ln{\left|\frac{L_{\perp}}{x-x'}\right|}, & 
\end{align}
\noindent where the correlation volume $v_0\sim \lambda^3$ and the integration over the 
variables $y'$ and $z'$ was restricted by the transverse size of the device, $L_{\perp}$ 
which is typically about few mm. The contribution of correlations is about 
$(\lambda /r_s)^3$ and remains much less than unity for all particle densities 
considered, therefore it does not affect the final estimation in 
Eq.~(\ref{energy-variation}).

For very low concentrations with $r_s\gg L$, assuming constant $p_s(x)\simeq r_s^{-3}$
all over the device one finds energy variation of the order of 
\begin{equation}
\label{variation1}
<\left( U_M(\mathbf{r},t)-<U_M(\mathbf{r},t)>\right)^2>\simeq 
\left( \frac{e^2}{4\pi\epsilon_0 \epsilon_s}\right)^2
\frac{4\pi L}{r_s^{3}}\ln{\frac{L_{\perp}}{L}}, 
\end{equation}
which cannot match the second term of Eq.~(\ref{micro-energy}) squared
at any reasonable value of $x_m<L$. This means domination of SP contribution for
these concentrations in accordance with the mean-force analysis in the previous
section.

For intermediate concentrations such that $l_T\ll r_s\ll L$ the equilibrium particle 
density may be used. Substituting Eq.~(\ref{macro-density}) into 
Eq.~(\ref{energy-variation}) and comparing the result with the second term of 
Eq.~(\ref{micro-energy}) squared one comes to the criterion
\begin{equation}
\label{SP-criterion}
x_m \simeq r_s \left(\frac{l_T}{16\pi r_s}\right)^{1/4}
\frac{1}{ \sqrt{ \ln{L_{\perp}/\lambda } } } 
\simeq (0.05 {-} 0.1) r_s 
\end{equation}
with the coefficient for $r_s$ by factor 0.25 less than that in the mean-force 
criterion~(\ref{MF-criterion}). When $x_m$ is less than the value given by the 
criterion~(\ref{SP-criterion}) the SP contribution dominates the injection process, in 
the opposite case the SP term becomes negligible comparing to the stochastic contribution
to the energy so that the MF mechanism of injection prevails.

For high particle concentrations with $r_s \lesssim l_T$, assuming again constant 
$p_s(x)\simeq l_T^{-3}$ the energy variation of the form (\ref{variation1}) obtains. 
In the considered parameter range it dominates strongly over the SP term for 
all reasonable values of $x_m$ so that the MF contribution prevails in accordance with 
the mean-force analysis in the previous section.

Summarizing analysis using both the mean-force and energy fluctuation criteria it is
apparent that the SP approach to the injection is in any case relevant for low 
concentrations of injected particles $p_s\leq  L^{-3}$ while the MF approach is valid 
for high concentrations $p_s\geq  l_T^{-3}$ whereas individual contributions of single 
particles and their images are lost in the collective mean field. 
In the intermediate range  $L^{-3} \leq  p_s\leq  l_T^{-3}$
the two mentioned criteria bring about somewhat different estimations of the crossover
from the SP to the MF domination regime, Eqs.~(\ref{MF-criterion}) and 
(\ref{SP-criterion}), that can be roughly compromised by the criterion
\begin{equation}
\label{criterion}
x_m \simeq 0.2 r_s,
\end{equation}
which will be used in the following analysis. The above specified microscopic conditions 
imply restrictions on the macroscopic variables and system parameters which define 
description of injection in terms of the SP or the MF approach when $x_m$ is smaller or 
larger than the value given by Eq.~(\ref{criterion}), respectively. The 
corresponding macroscopic conditions for fields and voltages are revealed in the next 
section.

\subsection{\label{subsec:Chart} Macroscopic criterion for the validity of the 
mean-field and the single-particle concepts}

\begin{figure}[!tbp]
\begin{center}
    \includegraphics[width=8cm]{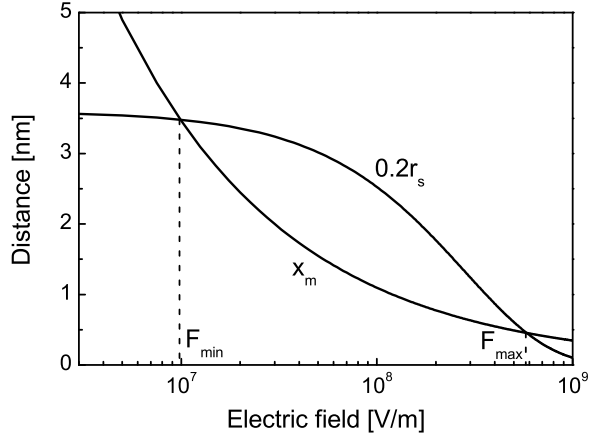}
    \caption{Position of the potential maximum of the Schottky barrier, $x_{m}$, 
             and $0.2$ of the mean distance between injected charge carriers in the MF 
             approach, $r_{s}$, as functions of the electric field at the interface for 
             $\Delta=0.225\mbox{ eV}$ and $T=300\mbox{ K}$.}
\label{x3Dr1D}
\end{center}
\end{figure}

\begin{figure}[!tbp]
\begin{center}
    \includegraphics[width=8cm]{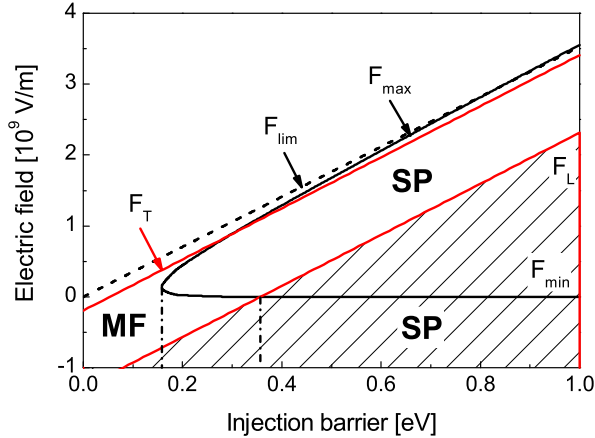}
    \caption{(Color online) Two field solutions of the equation $x_{m}=0.2r_{s}$ are shown with solid
    lines which delineate the areas where the MF and the SP concepts are 
    valid. The dashed line $F_{lim}$ restricts from above the field values calculated 
    under the assumption of Boltzmann statistics. 
    At the solid lines $F_{T}$ and $F_{L}$ the equations $r_{s}=l_T$ and $r_{s}=L$ are, 
    respectively, satisfied. Thereby the area above $F_{T}$ is described by the MF 
    approximation. In the hatched area below $F_{L}$ the continuous description of
    the transport and injection fails. To the left of the dash-dotted line at 
    $\Delta=0.16 \mbox{ eV}$ the inequality $0.2 r_{s}<x_{m}$ holds always. The other dash-dotted line at
    $\Delta=0.38 \mbox{ eV}$ marks the intersection of the lines $F_{min}$ and $F_{L}$.} 
\label{FrMaxMin}
\end{center}
\end{figure}

Let us consider the intermediate range of concentrations of injected charge carriers,
$L^{-3} \leq  p_s(0)\leq  l_T^{-3}$. Substituting expressions for 
$r_{s}=p_s(0)^{-1/3}$, Eq.~(\ref{boundary-cond-single}), and for $x_{m}$, 
Eq.~(\ref{x3D}), into the criterion~(\ref{criterion}) we arrive at a 
transcendent equation for the field $F_s(0)$ 
\begin{equation}
\label{x3D=r1D}
\sqrt{\frac{e}{16\pi\epsilon_0\epsilon_s F_s(0) }}= 0.2 
N^{-1/3}\exp{\left[ \frac{\Delta }{3kT} - \frac{eF_s(0)l_{TF}}{3kT}
\frac{\epsilon_s }{\epsilon_c} \right]}.
\end{equation}
\noindent A graphical solution of the above equation for typical parameters of ITO and 
organic semiconductors at room temperature (see Table~\ref{Materialparameters1}) is 
shown in Fig.~\ref{x3Dr1D} which exhibits two points of intersections. 
The dependence of the two corresponding field magnitudes on the barrier height $\Delta$ 
is shown in Fig.~\ref{FrMaxMin}. 
\noindent The MF approximation is expected to be valid in the range of interface fields 
smaller than $F_{min}$ or larger than $F_{max}$, where $0.2 r_{s}<x_{m}$, with the SP 
approximation prevailing between these lines  as indicated in Fig.~\ref{FrMaxMin} with 
account of the assumed restrictions $l_T < r_s < L$. 

The area of high carrier densities $p_s(0)\geq  l_T^{-3}$ is indicated by the line 
$F_{T}$ where the condition $r_{s}=l_T$ is fulfilled. Above this line determined by 
temperature the density of injected particles is so high that individual image forces 
are strongly overwhelmed by the mean field. Note also the line $F_{lim}$ indicating the 
border at which the Boltzmann statistics used in the MF consideration is violated. 

The preceding analysis in this section implied treatment in terms of the continuous 
particle density and mean field. For the low carrier densities $p_s(0)\leq  L^{-3}$ 
these quantities make no sense. The line $F_{L}$ indicates the border where the condition 
$r_{s}=L$ is fulfilled. Below this line determined by the device thickness the injected 
particles are so sparse that the continuous description fails but the SP approach 
applies. This hatched area in Fig.~\ref{FrMaxMin} overlaps with the above mentioned 
SP area but has another physical meaning. Here the Richardson-Schottky formula~\cite{Sze}
for the SP thermionic emission should apply or more sophisticated concepts adapted to the 
hopping conductivity typical of organic 
systems~\cite{Arkhipov1998,Arkhipov1999,Arkhipov2003,Reynaert2004,Woudenbergh2005} be 
implemented. In the upper SP region above the lines $F_{min}$ and $F_{L}$ the continuous 
description still works but the injection is dominated by the SP mechanism.

To the left of the characteristic value of the injection barrier of 
$\Delta \simeq 0.16\mbox{ eV}$, where the lines $F_{min}$ and $F_{max}$ merge, the MF 
approach is valid for all fields  between the values $F_{L}$ and $F_{lim}$. To the right 
of the characteristic value of the barrier of $\Delta \simeq 0.38 \mbox{ eV}$, at which 
the lines $F_{min}$ and $F_{L}$ intersect, the SP approach is valid for all fields up to 
the value of $F_{T}$. In the region of barrier values 
$0.16\mbox{ eV}<\Delta<0.38\mbox{ eV}$ the chart becomes nontrivial. Namely, by 
increasing the electric field from the very low values, firstly the only SP approach is 
applicable, then MF approach becomes applicable too and describes injection, later the 
reentrant SP regime prevails due to domination of individual image forces, and by further 
field increasing the reentrant MF regime is realized at high carrier concentrations when 
reaching the lower one of the fields $F_T$ or $F_{max}$. The drift-diffusion 
equation~(\ref{organic-eq}) applies thereby for the description of charge transport 
inside the organic semiconductor in the whole region $F_L < F_s < F_{lim}$.

Note that all the characteristic field lines presented in the chart on 
Fig.~\ref{FrMaxMin} result from estimations like the criterion (\ref{criterion}) 
and, hence, are not sharp boundaries but rather indicate the location of transition 
regions between the SP and MF domains. From the derivation of the 
criteria~(\ref{MF-criterion}), (\ref{SP-criterion}) and (\ref{criterion}) it is clear that
the uncertainty of the numerical factor in Eq.~(\ref{x3D=r1D}) may cause only a week logarithmic change in the 
characteristic fields.
The consequent uncertainty of the positions of the above field lines characterizes the 
width of the transition regions which appears to be much less on the logarithmic scale 
than the SP and MF regions themselves.

Now, using the obtained threshold values of the fields in Fig.~\ref{FrMaxMin}, we have 
to establish the range of voltages and barrier heights where the relevant field strengths 
can be achieved. Note that changing the injection barrier $\Delta$ for a fixed electrode 
means considering organic materials with different positions of the HOMO-level.

\begin{figure}[!tbp]
\begin{center}
    \includegraphics[width=8cm]{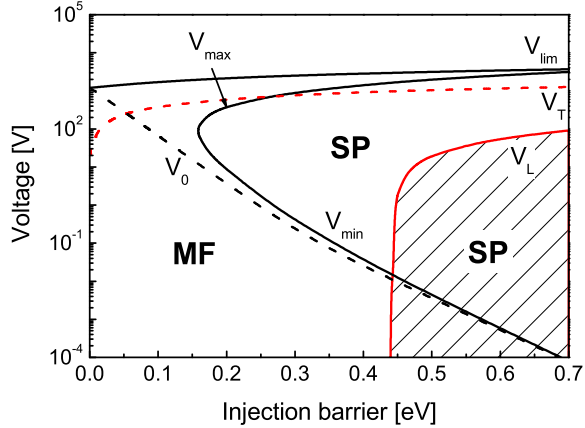}
    \caption{(Color online) Two solid lines, $V_{min}$ and $V_{max}$, outline areas of validity of 
             the SP and the MF approaches. The increasing dashed line $V_T$ further
             bounds the SP region from above. The line $V_{lim}$ bounds from above 
             the region where the Boltzmann statistics applies. The decreasing dashed 
             line $V_0$, shows the voltage at which the field at the interface vanishes. 
             The line $V_{L}$ delineates the boundary of the hatched area where the 
             mean field treatment loses any sense. } 
\label{VDelta-Chart}
\end{center}
\end{figure}

Combining the solution (\ref{solspecial}) with the boundary condition 
(\ref{boundary-cond-single}) one finds that the field at the interface is negative 
in equilibrium $(j=0)$ or when the applied voltage is small 
\cite{Cech2000,Neumann2006,Neumann2007}. The reason for this is that the field has to 
compensate the positive diffusion current caused by the huge difference in charge-carrier 
concentrations between the different sides of the interface. When a positive bias is 
applied, the field changes its sign at some point inside the dielectric usually called 
a "virtual electrode" \cite{Cech2000}. In the parameter range where the electric field 
is negative at the interface the Schottky barrier concept does not apply because a 
single charge carrier cannot escape from the potential well. In this case, however, the 
collective injection can occur which thus could be described within the MF approach. The 
boundary of this parameter region is delineated by the condition that the field at the 
interface equals zero, or, in other words, that the virtual electrode coincides with 
the physical interface. For a given value of the barrier the current magnitude at which 
this occurs may be found exactly using the solution (\ref{solspecial}) and reads 
\begin{equation}
\label{virtual_current}
j_0 = e \mu _s N \sqrt{ \frac{kTN}{2\epsilon_0\epsilon_s } }
|z_0|^{-3/2} \exp{\left( -\frac{3\Delta }{2kT} \right)},
\end{equation}   
\noindent where $z_0\simeq -1.02$ is the first zero of the Airy function 
$\mathrm{Ai}^{\prime}(z)$ (Ref.~\cite{Abramowitz}). The current-voltage (I-V) relation for a 
single ITO/organic interface is~\cite{Neumann2007}
\begin{align}
V & = l_{TF} \left[
\frac{\epsilon_s}{\epsilon_c} F_s(0) - \frac{j}{\sigma_c} \right]
+ 2\frac{kT}{e} 
\ln{\left[ 1- \frac{eL}{2kT} F_{s0}(0) \right]}\nonumber \\
& - l_{TF} \frac{\epsilon_s}{\epsilon_c} F_{s0}(0) 
- \frac{2kT}{e} \ln{
\left|
\frac{\mathrm{Ai}\left[\iota^{1/3}(L/l_{TF}+C_1)\right]}
{\mathrm{Ai}\left[\iota^{1/3}C_1\right]}
\right|},
\label{voltinf}
\end{align}
\noindent where the constant $F_{s0}(0)$ has to be determined from the boundary 
condition~(\ref{boundary-cond-single}) at the interface in the equilibrium case $j =0$
while the constants $F_{s}(0)$ and $C_1$ have to be determined satisfying 
Eq.~(\ref{boundary-cond-single}) with the solution (\ref{solspecial}).  
Using the current value $j_0$ and the current-voltage relation (\ref{voltinf}) 
a magnitude of the voltage $V_0(\Delta)$ may be obtained which 
defines the line $F_s(0)=0$ on the $V-\Delta $ plane (see Fig.~\ref{VDelta-Chart}). 
Using the characteristic field values of the Fig.~\ref{FrMaxMin}, also the voltages 
$V_{min},V_{max},V_{lim},V_T$ and $V_L$ which separate the MF and SP regions can be 
found by adjusting the current until the respective field values 
$F_{min},F_{max},F_{lim},F_T$ and $F_L$ are achieved at the interface as is presented 
in Fig.~\ref{VDelta-Chart}. The above calculations using continuous particle densities 
are applicable as long as the inequality $r_s<L$ roughly holds that is above the line 
$V_{L}$ below which only SP consideration makes sense. The line $V_{T}$ crosses the 
line $V_{max}$ so that the SP dominated region is bound from above by the voltage 
$\min{\{V_{max},V_{T}\}}$. Above this voltage the MF approach is valid until the 
voltage $V_{lim}$ is achieved at which it fails together with the Boltzmann statistics.

Though it looks counterintuitive, the continuous MF description of injection may be valid 
for low voltages while the discrete SP description, expected to work at low injection 
rates, applies also for rather high voltages. The reason can be seen in the very small 
escape distances $x_{m}$ at high voltages for which $x_m \ll r_s$, if the injection 
barrier $\bigtriangleup$ is not low, so that the individual image force on the charge 
carrier dynamics dominates. At low voltages and low barriers, however, the density of 
injected particles is diffusion-driven and therefore relatively high which reduces the 
interparticle distance $r_{s}$ to values small compared to the distance for the 
individual particle escape $x_{m}$ and entails domination of the mean field.

\section{\label{sec:modified-1electrode}Modified mean-field boundary conditions
                for a single interface}

The deficiency of the MF approach to the description of the injection within the SP 
parameter range in Fig.~\ref{VDelta-Chart} below the lines $\min{\{V_{max},V_{T}\}}$ 
and above the lines $\max{\{V_{min},V_{L}\}}$  is the missed contribution from the 
individual image forces of each single charge carrier. This strong but short-range 
deviation from the mean field near the interface may be accounted for as a dipole layer 
of a characteristic width $x_{m}$ as long as the latter distance is much less than 
$r_{s}$, which is true in the mentioned region. Indeed, this layer contributes to the 
shift in energy of each injected charge carrier through the work performed by the 
individual image force, which results in the shift in the electrostatic potential at the 
interface given by $\delta \phi_{Sch}$, Eq.~(\ref{Schottky-lowering}). Accounting for 
that in the mentioned SP region the boundary condition (requirement 1. of continuity of 
the electrostatic potential in the Section~\ref{subsec:density}) has to be replaced by 
the condition
\begin{equation}
\label{potential-jump}
\phi _s(+0) - \phi _c(-0) = \delta \phi_{Sch}. 
\end{equation}
Applying this boundary condition at $x=0$ the dipole layer is formally considered as 
an infinitesimal sheet so that the drift-diffusion equation (\ref{organic-eq}) applies 
for $x>0$.   

Taking this into account the modified boundary condition reads 
\begin{align}
\label{boundary-cond-3D}
\frac{p_{s}(0)}{N}&=\exp\left[ -\frac{\Delta }{kT} + 
\frac{eF_s(0)l_{TF}}{kT} \frac{\epsilon_s }{\epsilon_c}\right.\\\nonumber
&\left.
+\left( 1 - \frac{b}{0.2 r_s}  \right)
\frac{1}{kT}\sqrt{\frac{e^3 F_s(0)}{4\pi\epsilon_0\epsilon_s }}
\right]
\end{align}       
\noindent 
in the SP region restricted by the inequality 
$\max\{V_{min},V_L\}\leq V \leq \min\{V_{max},V_T\}$ with $b=\max\{x_m,0.2 l_T \}$. 
An interpolation factor $(1-b/0.2 r_s)$ introduced in the last term provides 
smooth switching of the dipole layer contribution at the boundary of the SP region 
which ensures that the position of this boundary is defined consistently when 
evaluated from inside and outside of this region. Indeed, crossing the line $V_{min}$
or $V_{max}$ from the SP region to the MF region the length $x_{m}$ exceeds 
$0.2 r_{s}$ and can no more be considered as the dipole layer thickness. In this case, 
the image force averaged over all injected carriers and their images is accounted for 
in the mean field value $F_s(0)$. The same occurs also at the line $F_T$ according to
analysis in Section~\ref{subsec:criteria}. Thus, the last term in the brackets in 
Eq.~(\ref{boundary-cond-3D}) following from the potential difference 
(\ref{potential-jump}), has to vanish at the boundary of SP region which is provided
by the above introduced interpolation factor.

 Now Fig.~\ref{VDelta-Chart} has to be reconsidered on the basis of the modified 
boundary condition (\ref{boundary-cond-3D}). It is obvious that the lines 
$V_{0},V_{min},V_{max}$ and $V_T$ do not change their positions since, at these lines, 
the boundary condition (\ref{boundary-cond-single}) still holds.
The equation $r_{s}=L$, however, is affected by the modified density of injected 
particles in Eq.~(\ref{boundary-cond-3D}). This is followed by the revised value of the
field $F_{L}$ producing the new borderline $V_L$. The MF approach applies now also in 
the SP region above the line $V_{L}$, however, with account of the dipole layer in 
Eq.~(\ref{boundary-cond-3D}), therefore it will be called from now on a modified 
mean-field (MMF) region. The genuine SP domain where the SP concept only applies shrinks 
now to the region below the $V_{L}$ line.
\begin{figure}[!tbp]
\begin{center}
    \includegraphics[width=8cm]{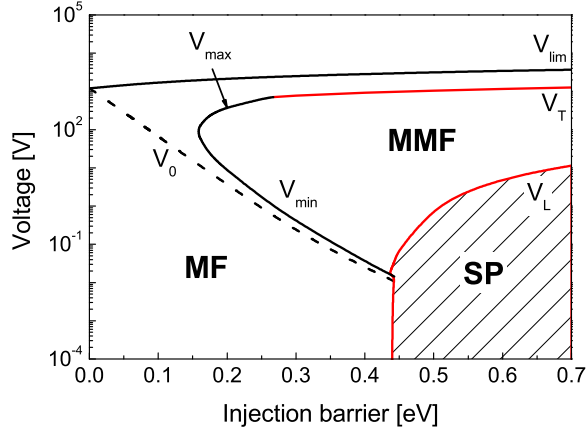}
    \caption{(Color online) Solid voltage lines $V_{min},V_{max}, V_{T}$ and, partly, $V_{L}$ outline the
             MMF area where the MF concept with the boundary condition (\ref{boundary-cond-3D}) 
             applies in contrast to the MF area where the boundary condition 
             (\ref{boundary-cond-single}) holds above and below the line $V_0$. The latter line
             shows the voltage at which the field at the interface vanishes. The MF approach
             fails completely only in the hatched SP region under the line $V_L$. The line 
             $V_{lim}$ bounds from above the region where the Boltzmann statistics applies.}
\label{VDelta-mod}
\end{center}
\end{figure}

Finally, the generalized boundary conditions (\ref{boundary-cond-single}) and
(\ref{boundary-cond-3D}) apply on the whole $V-\Delta$ plane below the $V_{lim}$ and 
above the $V_{L}$ lines and can be used together with standard continuous equations for 
the field and charge densities while the RS or FN single-particle concepts are expected 
to hold below the line $V_{L}$. To make these results comparable with experiment we 
address below two examples of practically relevant two-electrode devices.

\section{\label{sec:modified-2electrodes}Modified mean-field approach
                for two-electrode devices}

In the case of the organic layer sandwiched between two electrodes, located at the 
positions $x=\pm L/2$, multiple images of an injected hole will contribute to the 
potential in the SP approach. Consequently, the second term in Eq.~(\ref{single-barrier}) 
has to be replaced by the energy of charge interaction with both electrodes derived in 
the Appendix. We have to account now for possible injection of charge carriers from both 
electrodes. If the mean field $F_s$ becomes positive at the left interface at $x=-L/2$
or negative at the right interface $x=L/2$ (which never occurs at the same 
time~\cite{Neumann2006}) the SP Schottky mechanism of injection may become relevant at the 
respective electrodes. In the spirit of our modification of the boundary condition at 
the single interface $x=0$ in Eq.~(\ref{potential-jump})-(\ref{boundary-cond-3D}), the boundary 
conditions at $x=\pm L/2$ should be modified by the potential discontinuities 
$\delta \phi_{Sch}^{\pm}$ if the distance from the potential maximum to the respective 
electrode $L/2-|x_m|<0.2 r_s^{\pm}$ with $r_s^{\pm}=\left[p_s (\pm L/2) \right]^{-1/3}$
being the mean distance between injected particles near the electrodes at $\pm L/2$. 
In this case, the mean field may be assumed constant within the distances $r_s^{\pm}$ 
from the respective electrodes and the following approximations used for the energy 
profile near the electrodes    
\begin{align}
 U_{\pm} \left(x \right) & \cong \Delta^{\pm} - eF_s 
\left(\pm L/2 \right)\left(x\mp L/2 \right) \nonumber\\ 
&+ \frac{e^2}{16 \pi \epsilon_0 \epsilon_s L} \left[ \psi \left( \frac{1}{2} 
+ \frac{x}{L} \right) + 
\psi \left( \frac{1}{2} - \frac{x}{L} \right) + 2\gamma  \right],
\label{2electrode-profile} 
\end{align}  
\noindent where $\Delta^{\pm}$ denote the injection barriers at $x=\pm L/2$, 
$\gamma$ is Euler's constant and $\psi(x)$ is the digamma function~\cite{Abramowitz}. 
The expression in square brackets reproduces the behavior of the Coulomb potential in 
Eq.~(\ref{single-barrier}) when approaching one of the electrodes at $x=\pm L/2$ but is 
symmetric and has a maximum at $x=0$. Note that the criterion (\ref{criterion}) derived 
using the energy (\ref{single-barrier}) for a single interface accounts, in fact, for 
the second electrode and can therefore be applied to the two-electrode devices.

According to the approximations (\ref{2electrode-profile}) the position of the maximum 
is defined by one of the equations
\begin{equation}
\label{maxima}
\psi^{\prime} \left( \frac{1}{2} + \frac{x_m}{L} \right) -
\psi^{\prime} \left( \frac{1}{2} - \frac{x_m}{L} \right) =
\frac{16 \pi \epsilon_0 \epsilon_s L^2}{e} F_s \left( \pm L/2 \right),
\end{equation}  
\noindent taking the upper sign, if $F_s \left(L/2\right)<0$, or the lower sign, if 
$F_s \left(-L/2\right)>0$. The respective potential discontinuities are given 
then by the formula 
\begin{equation}
\label{jumps}
e\delta \phi_{Sch}^{\pm}=\Delta^{\pm} - U_{\pm} \left(x_m \right),  
\end{equation} 
\noindent which determines them as functions of the field at the respective 
interfaces.

The boundary conditions 1.-3. of Section \ref{subsec:density} modified with the 
potential discontinuities (\ref{potential-jump}) and (\ref{jumps}) at both interfaces 
can now be written as
\begin{align}
\label{boundary-MF-sym}
\frac{p_{s}(\pm L/2 )}{N}  &= \exp\left[ 
-\frac{\Delta^{\pm} }{kT} \mp 
\frac{eF_s(\pm L/2 )l_{TF}^{\pm}}{kT}  \frac{\epsilon_s }{\epsilon_c^{\pm}}    
\right.\nonumber\\
&\left. + \left( 1 - \frac{b^{\pm}}{0.2 r^{\pm}_s}  
\right)\frac{e\delta \phi_{Sch}^{\pm} }{kT}
\right],
\end{align}
\noindent where quantities identified with superscript $\pm$ denote here and below 
the parameters of the two electrodes contacted at $x=\pm L/2$. Note that the last term
in the exponent appears only inside the MMF area on the V-$\Delta$ plane whose boundaries
are determined self-consistently by the criterion $L/2-|x_m|=0.2 r_s^{\pm}$ using the
formula (\ref{boundary-MF-sym}) itself. The cases of symmetric and asymmetric devices 
are considered separately below.

\subsection{\label{subsec:2electrodes-symmetric}Modified validity chart and 
current-voltage characteristics of a unipolar symmetric organic device}

We examine first the case of a symmetric unipolar device consisting of two 
ITO-electrodes and an organic layer in between providing $p$-type conductivity. At zero 
bias the diffusion-mediated electric field is directed outwards at both electrodes, i.e. 
is negative at $x=-L/2$ and positive at $x=L/2$. In this case the Schottky-type 
contribution in the formula (\ref{boundary-MF-sym}) vanishes and the MF boundary 
conditions of the type of Eq.~(\ref{boundary-cond-single}) apply at both interfaces 
as long as the barrier is not too large which is in agreement with the chart for the 
single ITO/organic interface (Fig.~\ref{VDelta-mod}). Note, further, that for a positive 
bias, a positive field always prevails at the right, collecting electrode so that the 
Schottky-type barrier lowering does not arise here at all and this electrode remains in 
the MF regime as long as the continuous approach is valid.

On the other hand, when the bias is high enough, the field at the left, injecting 
electrode becomes positive and may exceed the characteristic field $F_{min}$ in
which case the modified boundary condition (\ref{boundary-MF-sym}) includes the
Schottky lowering term. Since $\delta\phi_{Sch}^{-}$ is not known as explicit function 
of the field $F_s(-L/2)$ the equation (\ref{boundary-MF-sym}) should be solved together 
with Eqs.~(\ref{2electrode-profile})-(\ref{jumps}). Then the mean interparticle distance 
$r_s^{-}$ can be evaluated and the equation $L/2-|x_m|=0.2 r_s^{-}$ solved with 
respect to the field $F_s(-L/2)$ resulting in the positions of the lines $F_{min}$ and 
$F_{max}$ on the $F-\Delta$ plane for the injecting electrode. Additionally, the 
characteristic field $F_{lim}$ can also be found equating the exponent in 
Eq.~(\ref{boundary-MF-sym}) to zero, and the field  $F_T$ can be determined from the 
equation  $r_s^{-}=l_T$. Determination of the line $F_L$, where the continuous approach 
fails, is a bit less straightforward. When the charge carriers are so sparse in the 
device that the mean distance between them equals the distance between the electrodes 
their concentration is defined by injection from both electrodes. Therefore the mean 
particle density $\bar p_s$ and the corresponding mean distance 
$\bar r_s=\bar p_s^{-1/3}$ should be used for the definition of $F_L$ from equation 
$\bar r_s=L$. Integrating the Gauss equation over the device thickness one finds that 
\begin{equation}
\label{mean-ps}
\bar p_s = \frac{\epsilon_0\epsilon_s}{eL}\left[ F_s(L/2)-F_s(-L/2) \right].
\end{equation}
Thus the procedure of $F_L$ determination is as follows. Below the line $F_{min}$ the
MF boundary conditions, Eq.~(\ref{boundary-cond-single}), are satisfied at both 
electrodes with Eq.~(\ref{solspecial}) and the current is increased until equation 
$\bar r_s=L$ is fulfilled. Above the line $F_{min}$ the boundary 
condition~(\ref{boundary-MF-sym}) should be satisfied at the left electrode and the
MF boundary condition~(\ref{boundary-cond-single}) at the right electrode while the
current is adjusted until the requirement $\bar r_s=L$ is fulfilled. The mentioned lines 
delineate together the MMF region on the field-barrier plane similar to 
Fig.~\ref{FrMaxMin}. This chart, however, plays only an auxiliary role and will not 
be presented here.

Having determined $F_{min}$, $F_{max}$, $F_{lim}$, $F_L$ and $F_T$  as functions of 
$\Delta^{-}$ one can calculate, using the solution~(\ref{solspecial}), the respective 
voltages $V_{0}$, $V_{min}$, $V_{max}$, $V_{lim}$, $V_L$ and $V_T$ which delineate the 
MMF region. To this end, the current is varied as parameter until the respective field 
magnitude is achieved at $x=-L/2$; the voltage then follows by direct integration of the 
field over the device thickness~\cite{Neumann2006,Yampolskii2008}. Thereby the free 
constants $C_1$ and $C_2$ in Eq.~(\ref{solspecial}) are obtained by satisfying the 
conditions~(\ref{boundary-MF-sym}) and (\ref{boundary-cond-single}) at $x=-L/2$ 
and $x=L/2$, respectively. This method, however, appears to be numerically unstable 
because of fast oscillations of the Airy functions providing multiple solutions for the 
constants. Alternatively, Eq.~(\ref{organic-eq}) can be directly numerically integrated 
using the same boundary conditions which proves to be a robust procedure.

\begin{figure}[!tbp]
\begin{center}
    \includegraphics[width=8cm]{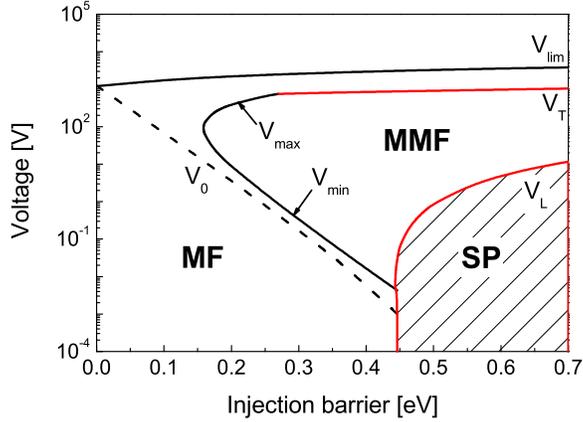}
    \caption{(Color online) Solid lines $V_{min}, V_{max}, V_{T},$ and, partly, $V_{L}$ outline the MMF 
             area for a symmetric ITO/organic/ITO device where the SP contribution to the 
             MF approach must be accounted for. The dashed line indicates the voltages 
             $V_{0}$ at which the field at the left interface vanishes. The line 
             $V_{lim}$ bounds from above the area where the Boltzmann statistics applies
             and the line $V_L$ the hatched area where the continuous description fails 
             completely.}  
\label{chartITOsym}
\end{center}
\end{figure}

The resulting chart on the $V-\Delta^{-}$ plane is depicted in 
Fig.~\ref{chartITOsym}. 
\noindent Apparently, this chart is very similar to that of the single interface 
depicted in Fig.~\ref{VDelta-mod} except the voltage region below 0.1 V where the 
$V_{0}$ and $V_{min}$ curves go down a bit steeper. This means, on the one hand, that 
the injecting interface dominates the behavior of a symmetric device. On the other hand, 
this similarity shows that accounting for the collecting electrode in the energy profile 
of a single particle, Eq.~(\ref{2electrode-profile}), does not add much to the simpler 
formula for a single interface, Eq.~(\ref{single-barrier}).

\begin{figure}[!tbp]
\begin{center}
    \includegraphics[width=8cm]{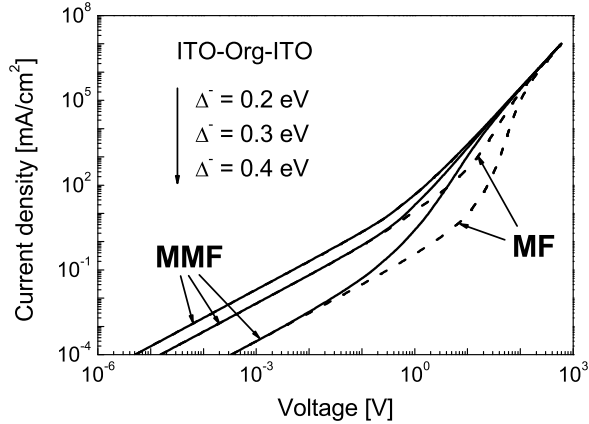}
    \caption{I-V characteristics for symmetric organic devices with and without account 
             of charge carrier discreteness are shown by solid and dashed lines, 
             respectively. The barrier magnitude is changed from the upper to the lower 
             curve as is shown in the plot. For the barrier values below 0.2 eV and 
             smaller the solid and the dashed lines cannot be distinguished.} 
\label{CV-sym}
\end{center}
\end{figure}

I-V characteristics for different injecting barrier values are plotted in 
Fig.~\ref{CV-sym}  
\noindent for a symmetric device, employing typical parameters of ITO and organic 
semiconductors (see Table~\ref{Materialparameters1}). The essential contribution of the 
SP Schottky barrier lowering in the wide region of voltages is evident.

\subsection{\label{subsec:2electrodes-asymmetric}Modified validity chart and 
                   current-voltage characteristics of an unipolar asymmetric 
                   organic device}

Asymmetric devices are considerably different from symmetric ones due to the presence of 
the built-in potential, $V_{bi}$, which is defined by the difference between the work 
functions of the two electrodes, 
$eV_{bi} = E_A^{+}-E_A^{-}=E_b^{+}+\kappa _{\infty}^{+}-E_b^{-}-\kappa _{\infty}^{-}$. 
At equilibrium, due to equalization of the electrochemical potential in the entire 
system, an internal electric field emerges. For that reason the electric field at one of 
the electrodes may change its sign providing favorable conditions for the SP injection 
scenario already at zero bias. To account for this effect the field-barrier charts should 
be considered separately for both electrodes by solving equations   
$L/2-|x_m|=0.2 r_s^{\pm}$ which result in corresponding characteristic fields
for both electrodes. Consequently, 
Eq.~(\ref{organic-eq}) should be solved implementing the boundary
conditions~(\ref{boundary-MF-sym}) self-consistently at $x=\pm L/2$. Thereby potential 
discontinuities $\delta\phi_{Sch}^{\pm}$ from Eq.~(\ref{jumps}) should be applied in 
Eq.~(\ref{boundary-MF-sym}) as long as 
$\mbox{max}\{|F_{min}^{\pm}(\Delta^{\pm})|,|F_L^{\pm}(\Delta^{\pm})|\}<|F_s(\pm L/2)|
<\mbox{min}\{|F_{max}^{\pm}(\Delta^{\pm})|,|F_{lim}^{\pm}(\Delta^{\pm})|,|F_T^{\pm}(\Delta^{\pm})|\}$. 
Note that the fields $F_L^{\pm}$ are defined simultaneously from the mean value of the
carrier density using Eq.~(\ref{mean-ps}) as for the symmetric device. As soon as the 
characteristic fields are known the corresponding characteristic voltages may be 
evaluated as was described in the preceding section. The line $V_L$ is obviously common
for the $V-\Delta^{\pm} $ charts of the left and right electrodes.

Application of 
the self-consistent boundary conditions~(\ref{boundary-MF-sym}) may look 
a bit cumbersome but it is substantially simplified by the fact that the sole maximum of 
the energy~(\ref{2electrode-profile}), $x_m$, occurs at each moment only near one of the 
two electrodes. Thus, if the external bias has the same sign as the built-in potential 
then the SP contribution may arise at the injecting electrode only. If the external bias 
has the sign opposite to that of the built-in potential the SP contribution may arise 
first in equilibrium at the collecting electrode due to the internal electric field while 
the injecting electrode is described in the MF approach. By the increasing external 
field the collecting electrode goes over to the MF regime too and by further field 
increasing the injecting electrode goes over to the SP regime where the discontinuity 
$\delta\phi_{Sch}$ applies.

\begin{figure}[!tbp]
\begin{center}
    \includegraphics[width=8cm]{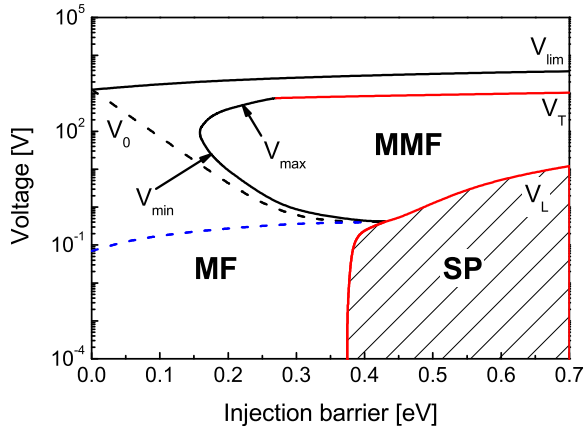}
    \caption{(Color online) Solid lines $V_{min}, V_{max}, V_{T}$ and, partly, $V_{L}$ outline the MMF 
             area for an asymmetric ITO/organic/Al device where the SP contribution to 
             the MF approach must be accounted for at the left, injecting electrode. The 
             upper dashed line indicates the voltages $V_{0}$ at which the field at the 
             left interface vanishes. The lower dashed line delimits the low-voltage 
             region, where the SP contribution to the MF approach must be accounted for 
             at the right, collecting Al electrode. In the hatched SP region bound by the 
             line $V_L$ the description of the device with the continuous carrier density 
             fails completely. The Boltzmann statistics is valid below the line 
             $V_{lim}$.}  
\label{chartITOasym}  
\end{center}
\end{figure}

To investigate possible effects induced by the presence of the built-in potential, 
the case of an asymmetric unipolar device consisted of the organic layer sandwiched 
between ITO and Al electrodes is considered. For the ITO electrode at $x=-L/2$ and 
Al electrode at $x=L/2$ the built-in potential equals -0.4 V (see the electrode 
parameters in Table~\ref{Materialparameters2}). 
\begin{table}
\caption{\label{Materialparameters2}Parameter values for collecting electrodes
(see Ref.~\onlinecite{Ashcroft}). }
\begin{ruledtabular}
\renewcommand{\arraystretch}{1.5}
\begin{tabular}{cccccc}
& $l_{TF}^+$ & $p_{\infty}^+ $ & $ \mu_c^+ $ & $ \kappa_{\infty}^+ $ & $E_A^+$ \vspace{2pt}\\
& ($\mathring{\text A}$) & (cm$^{-3}$) &  $\left( \displaystyle \frac{\text{cm}^2}{\text{V s}}\right)$ & (eV) & (eV) \vspace{2pt}
\\
\hline
Al & 0.5 & $1.81 \cdot 10^{23} $ & 13.05 &  11.7 & 4.3 \\
Au & 0.6 & $5.9 \cdot 10^{22} $ &  44  & 5.53 & 4.3 
\end{tabular}
\end{ruledtabular}
\vspace{0.5cm}
\end{table}
The $V-\Delta^{-}$ chart for this system is presented in Fig.~\ref{chartITOasym} where
the behavior of the right electrode is also comprised. 
Changing the injection barrier $\Delta^{-}$ for fixed electrodes means considering organic 
materials with different positions of the HOMO-level. This implies, in turn, the change of 
the injection barrier at the other electrode, 
$\Delta^{+}$, to the same extent. This allows one to show the area where the SP injection
mechanism is relevant at the right electrode on the same $V-\Delta^{-}$ chart.

The $V_0$ curve for the asymmetric case, where the field vanishes at the left, injecting 
interface, nearly coincides with that of the symmetric case (cf. Fig.~\ref{chartITOsym}) 
for barrier values below $\Delta^{-}=0.22\mbox{ eV}$. Above the value of $0.35\mbox{ eV}$ 
it coincides virtually with the lower boundary of the MMF region given by the $V_{min}$ 
curve because $F_{min}$ value becomes equal to zero. When $\Delta^{-} > 0.25\mbox{ eV}$, 
the $V_0$ curve differs substantially from the corresponding line for the symmetric 
device and saturates at about the value of $-V_{bi}$. The $V_{lim}$ curve which delineates 
the upper boundary of the chart almost coincides for the symmetric and asymmetric cases. 
Thus, the presence of the built-in potential reveals itself in the extension of the 
low-voltage area, where the Schottky correction can be neglected at the injecting 
electrode, up to approximately $-V_{bi}$. On the other hand, this correction should be
included at the collecting electrode at almost all voltages below $-V_{bi}$ as is indicated
by the lower dashed line.

\begin{figure}[!tbp]
\begin{center}
    \includegraphics[width=8cm]{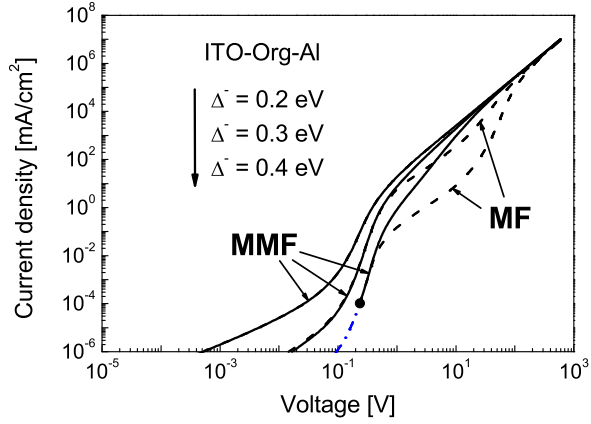}
    \caption{(Color online) I-V characteristics for asymmetric organic devices with and without
             account of charge carrier discreteness are shown by solid and dashed 
             lines, respectively. The barrier magnitude is changed from the upper to the
             lower curve as is indicated in the plot. The solid circle on the 
             characteristic for $\Delta = 0.4$~eV marks where it enters to the 
             low-voltage region (dash-dotted line) where the mean-field approach is 
              no more valid.} 
\label{CV-ITOAL}
\end{center}
\end{figure}

I-V characteristics for different injecting barrier values are plotted in 
Fig.~\ref{CV-ITOAL}   
for the asymmetric device, employing typical parameters of ITO and organic 
semiconductors \cite{Neumann2007}. As in the symmetric case, the Schottky lowering of 
the barrier enhances the current by orders of the magnitude for high voltages compared 
to the pure MF description~\cite{Neumann2007}. This results in the disappearing 
of the second exponential increase in the case of asymmetric devices and, generally, in 
an earlier transition to the space charge limited regime in both symmetric and 
asymmetric cases. Note that account for the charge carrier discreteness does not change 
the I-V curves below $-V_{bi}$ in Fig. \ref{CV-ITOAL}.

\section{\label{sec:experiment}Comparison with experiment}

To evaluate the presented model, the results for an asymmetric device are compared with 
experimental data obtained on a unipolar device consisting of a single 
poly[{\it p}-(2-methoxy-5-(3,7-dimethyloctyloxy)phenylene-vinylene] (OC$_1$C$_{10}$-PPV)
layer of thickness $L=100$~nm sandwiched between ITO and Au electrodes. The 
characteristic energy of the HOMO level of the employed OC$_1$C$_{10}$-PPV is 5~eV and the hole mobility 
$\mu_s$ is reported between $5 \times 10^{-7}$ and $ 3 \times 10^{-5}$~cm$^2$/(V~s) or even 
larger~\cite{Blom,Blom2000,Geens2002,Krebs2003}. Note that for an organic semiconductor 
the mobility is typically scattered by orders of the magnitude~\cite{deBoer2004}. 
It is known also that the work function of ITO is sensitive to the cleaning 
procedure and thus, it may vary from 
4.7 to 5~eV (up to the energy of the HOMO level). Additionally, a dipole layer may 
possibly emerge at the metal/organic interface, typical for these 
systems~\cite{Ishii,Kahn}. Therefore, we consider the injection barriers $\Delta^-$ and 
$\Delta^+$ as independent fitting parameters.

\begin{figure}[!tbp]
\begin{center}
    \includegraphics[width=8cm]{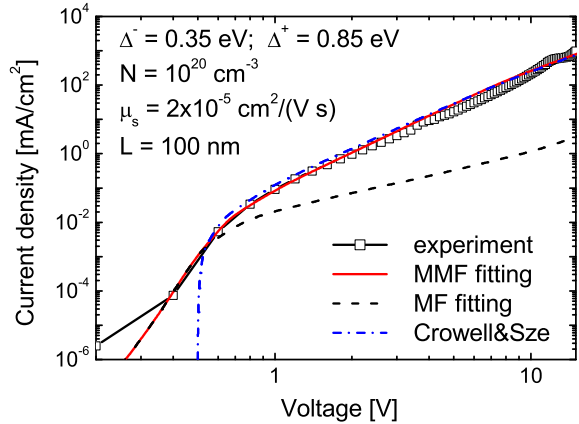}
    \caption{(Color online) Measured I-V characteristic of ITO/OC$_1$C$_{10}$-PPV/Au structure (squared curve)
and its approximations within different theoretical models. The best fitting parameters 
in the MMF approach are indicated in the plot.} 
\label{fitting}
\end{center}
\end{figure}

In Fig.~\ref{fitting} the measured I-V characteristic of the ITO/OC$_1$C$_{10}$-PPV/Au structure 
is shown as well as the I-V dependencies calculated for different model approximations.
One can see, that the advanced MMF approach gives a good approximation for the 
measured I-V curve in the wide range of applied voltages with the best fitting parameters 
indicated in the figure. At the same time, the dashed curve for the MF approach with the 
same fitting parameters underestimates the current by orders of the magnitude at high 
voltages. Note that the widely used Crowell and Sze model~\cite{Crowell1966,Blom2000} 
(the dash-dotted curve) fails to describe the I-V characteristic at voltages below 
$-V_{bi}$.

Notice, however, that the best fit was found for slightly higher magnitudes of the 
injection barriers than the largest values reported in the literature: 
$\Delta^-_{max}=0.3$~eV and $\Delta^+_{max}=0.7$~eV (the work function of Au is 4.3~eV, 
see Table~\ref{Materialparameters2}). This deviation of the barrier values may originate 
from the uncertain position of the HOMO level in OC$_1$C$_{10}$-PPV depending on details of the material 
production as well as from the above mentioned possible dipole layers at the electrode/OC$_1$C$_{10}$-PPV 
interfaces. Also, the {\it exact} shape of the I-V curve cannot be fully approximated 
because of the fact that the proposed model still misses some important features of 
insulators, in particular of organic semiconductors. Those are the realistic DOS shape 
including energetically distributed trap states~\cite{Lambert,Blom} and the field- and 
concentration-dependent carrier mobility~\cite{Blom2000,Pai} characteristic of organic 
semiconductors. Remarkable deviation of the theory from the experiment below the 
current densities of $10^{-4} \rm \: mA/cm^2$ may have origin in some other physical 
mechanisms as for example leakage effect or impurity doping. Nonetheless, a sufficiently 
good matching of experiment and MMF model is found over 2 orders of magnitude in the 
applied voltage and 7 orders of magnitude in the current density.

\section{\label{sec:conclusions}Discussion and Conclusions}

In this work, we have advanced boundary conditions for a continuous description of
charge injection and transport in dielectrics in terms of mean electric field and 
charge density accounting for discreteness of charge carriers. The latter becomes
important for a wide range of values of injection barriers and voltages applied
where individual image force dominates the injection process. Regions of parameters,
where charge carrier discreteness plays a decisive role, are approximately delineated
in the $V-\Delta$ charts for symmetric and asymmetric conductor/insulator/conductor 
devices using an 
organic semiconductor as example for an insulator sandwiched between two electrodes. 
Implementing generalized boundary conditions allows for the description of charge 
transport in such systems in both low injection and high injection regimes including space 
charge limited transport. The advanced theory applies to a wide class of devices
where injection of charge carriers into dielectrics takes place, particularly, to
organic light-emitting diodes, thin film ferroelectric capacitors, etc. 
Though in this work only the problems of simple plain geometry were treated, the 
developed approach is not intrinsically one-dimensional and can be applied as well to 
complicated two- and three-dimensional configurations of conductor/insulator interfaces 
as, for example, those of organic field-effect transistors, however, with duly adapted
boundary conditions (\ref{boundary-MF-sym}).

The presented description of injection applies directly to inorganic crystal insulators 
and wide-gap non-degenerate semiconductors as well as to the very narrow-band insulators and 
semiconductors as was indicated in Section \ref{subsec:density}. When applying this 
theory to the wide-gap organic semiconductors one should take into account specific 
features of these media. Being disordered semiconductors with typically hopping
conductivity, organic systems differ from inorganic crystals mostly in two respects 
concerning the injection process, namely, they 1) possess as a rule a relatively wide 
(Gauss) DOS distribution without sharp band 
edges~\cite{Mensfoort2008,Arkhipov1998,Arkhipov1999,%
Burin2000,Arkhipov2003,Woudenbergh2005,Holst2009,Pasveer2005} and 2) possibly exhibit 
long-range correlations of the random energy levels~\cite{Dunlap1996,Novikov1998}.

Considering point~1), the absence of a sharp band edge makes the injection barrier an 
ill-defined parameter. If the barrier $\Delta$ is defined as a difference between the 
Fermi level in the injecting electrode and the center of the density of states 
distribution at the HOMO-level as it was done in 
Refs.~\cite{Arkhipov1998,Arkhipov1999,Burin2000,Arkhipov2003} the injection into
the "gap" states below the barrier is thereby implied because of numerous states 
available at least to the depth of $\Delta-\sigma$, where $\sigma$ is the width of the 
Gauss DOS. Nevertheless, a nontrivial fact is the possible injection into the
tail states well below  the 
barrier~\cite{Arkhipov1998,Arkhipov1999,Burin2000,Arkhipov2003} (to the depth of about 
$\Delta - \sigma^2/kT$) which results in the field-dependent mobility of 
carriers~\cite{Burin2000} or in the allegedly possible space charge 
regime~\cite{Arkhipov2003} for high injection barriers. The problem of the concept of 
injection into the deep tail states is that the injection is considered as a 
single-particle process ignoring the fact that these states may be occupied. Particularly 
problematic becomes, therefore, a combination of the SP injection with the drift-diffusion 
equation~\cite{Arkhipov2003} at high injection levels because in the space-charge regime 
the density of injected charge is especially high near the injecting electrode. On the 
other hand, the concept of injection into the tail states may be well applicable at low 
injection levels, particularly, in the genuine SP region of the V-$\Delta$ chart of 
Fig.~\ref{chartITOasym}. In our simulation of the experiment in Fig.~\ref{fitting} we 
have assumed a narrow-band approximation for the HOMO-level ignoring the finite width of 
the DOS and, thus, the disorder effect on injection. This might explain the deviation 
of the theoretical characteristic from the experimental one at low voltages. In fact, 
to ignore the role of disorder on injection the inequality $\sigma \ll kT$ should be 
satisfied which is not typically the case for organic semiconductors. Proper integration 
of the finite DOS width in our self-consistent concept is currently in progress.

A rather elaborated many-particle description of charge injection and transport in 
organic semiconductors with uncorrelated Gauss disorder was developed recently in the 
Ref.~\cite{Holst2009} where boundary conditions similar to our MMF ones were used to 
account for the image charge effect. Two models were presented and compared in this 
paper, a discrete model of 3D hopping over the sites of a cubic lattice and a
continuous 1D description in the spirit of the drift-diffusion concept, which exhibit
good agreement with each other. However, in both models the individual image effect
was substantially overestimated. In the 3D approach the individual image 
contribution was accounted for at all internal sites of the lattice in addition to the
mean field derived from the charge averaged over the planes parallel to electrodes. As 
was discussed above in Sections \ref{sec:generalmodel} and
\ref{sec:modified-1electrode} the individual image effect reveals itself as strong 
deviation from the mean field only in the vicinity of the electrodes due to its 
short-range nature. In fact, close to the electrode surface the individual image force 
may dominate over the mean field. In contrast to this, far from the electrode a charge 
carrier "observes" not only its own image but all the other injected particles with their 
images which together constitute the mean field. In the latter case there is no reason to 
single out the carrier's own image which would mean duplication of the image account. 
This problem, was discussed in detail in Refs.~\cite{Bussac1998,Tutis1999,Tutis2001} 
where such
duplication was explicitely excluded. In the continuous 1D model of Ref.~\cite{Holst2009}, 
the line $V_0$ was chosen as a boundary between the MF and MMF region, where the barrier 
lowering due to the image effect is included. This line is close to our lower boundary 
of the MMF region $V_{min}$ on the V-$\Delta$ charts in Figs.~\ref{VDelta-mod}, 
\ref{chartITOsym} and \ref{chartITOasym} for a certain range of barrier values. 
This approach, however, also overestimates the image effect for low barriers and misses 
the fact that below some threshold barrier value of about 0.2 eV the MF approach is always 
valid independently of the field direction at the electrode.

The property~2) of organic semiconductors which may be relevant to statistical 
grounds of our concept of injection, namely, the correlated disorder of the energy levels 
at different sites is not so well established and is still disputed in the literature.    
The idea of the correlated disorder was suggested in Refs.~\cite{Dunlap1996,Novikov1998}
to explain the field dependence of mobility $\ln{\mu}\sim \sqrt{F}$ in a wide region of 
field values. We note, firstly, that this single-particle concept is based on the 
arbitrary hypothesis of independently and randomly oriented dipoles at each lattice 
site~\cite{Novikov1998}. Secondly, such field dependence of the mobility was also explained within
another single-particle model considering the injection in the deep tail states of the 
uncorrelated Gauss disordered energy levels~\cite{Burin2000}. Many-particle models 
assuming uncorrelated disorder combined with concentration dependence of the 
mobility~\cite{Holst2009,Pasveer2005} can provide good agreement with experiments, too. 
Pasveer et al. concluded that there is no necessity to account for correlated disorder
at least around room temperature~\cite{Pasveer2005}. Hence, possible correlations in 
charge positions due to the correlated disorder are not regarded in our study.

Considering the Schottky-barrier lowering we have to note the 
general problem of this concept caused by the too large distances between the electrode 
and the potential barrier maximum at low voltages. In fact, this value is restricted for 
different physical reasons depending on the material type. In crystalline dielectrics 
this concept, implying usually ballistic overcoming the potential barrier by a charge 
carrier~\cite{Sze}, makes sense as long as the mentioned distance does not exceed the mean 
free path which may be large enough. On the other hand, in organic semiconductors 
characterized by hopping conductivity the ballistic description fails because of typical 
hopping distance of $0.1 {-} 1$ nm. In this case, more sophisticated descriptions of injection 
like multiple random hopping of a single charge carrier in the SP potential may be 
relevant~\cite{Arkhipov1998,Arkhipov1999}. Applicability of the latter model seems to be 
restricted to the SP region in the charts of Figs.~\ref{VDelta-mod}, \ref{chartITOsym} 
and \ref{chartITOasym} where the effect of the other carriers is reduced. After all, we 
would like to stress that in the formulation of the thermodynamic boundary conditions 
1.-3. of Section~\ref{subsec:density} together with Eq.~(\ref{potential-jump}) the 
ballistic mechanism of charge injection was not assumed.

Finally, we considered in this paper the corrections to the self-consistent mean-field 
description of injection due to the effect of discreteness of charge carriers. It is 
apparent, however, that below the line $V_L$, where continuous description is no more 
valid, the pure SP approach will also be distorted because of the long-range interaction 
with the other injected charge carriers at the distance $L$ or larger. This is a sort of 
MF corrections to the SP picture which is yet to be elaborated.

\begin{acknowledgments}
Useful discussions with A. Klein and A. Tagantsev are gratefully acknowledged. 
This work was supported by the Deutsche 
Forschungsgemeinschaft through the Sonderforschungsbereich 595.
\end{acknowledgments}

\appendix*
\section{Energy of a point charge between two plane electrodes}

When a point charge $q$ is located in the dielectric medium with relative permittivity 
$\epsilon_s$ between two plane electrodes taking positions at $x=\pm L/2$, the Coulomb 
potential used in the formula~(\ref{single-barrier}) for the single interface does not 
satisfy anymore the boundary conditions for the electrostatic potential at the 
electrodes:~$\varphi (x=\pm L/2)=0$ (no voltage applied). To satisfy these boundary 
conditions the method of images may be used~\cite{LandauElectrodynamicsContinuum} which 
results, in the case of two electrodes~\cite{Genenko3}, in the potential expression
\begin{equation}
\label{Coulomb-2el}
\varphi(x,y,z) = \frac{q}{4 \pi \epsilon_0 \epsilon_s}
\sum_{n}\frac{(-1)^n}{\sqrt{(x-x_n)^2+y^2+z^2}},
\end{equation}
\noindent where $(x_0,0,0)$ is the position of the point charge and $x_n=x_0(-1)^n+nL$ 
with $n = \pm 1, \pm 2, \dotso$ are the $x-$coordinates of its images.

The force, exerted upon the point charge, results from the interaction with image charges
to the left and to the right of its position:
\begin{align}
\label{force-2el}
f(x_0)  & = \frac{q^2}{4 \pi \epsilon_0 \epsilon_s }
\sum_{n=1}^{\infty}(-1)^n \left[\frac{1}{(x_0-x_{-n})^2}-\frac{1}{(x_0-x_n)^2}\right]
\nonumber\\
&= \frac{2q^2x_0}{\pi \epsilon_0 \epsilon_s L^3 }
\sum_{m=0}^{\infty} \frac{2m+1}{\left[(2m+1)^2-(2x_0/L)^2\right]^2} 
\nonumber\\
&= \frac{q^2}{16 \pi \epsilon_0 \epsilon_s L^2}
\left[ \psi^{\prime}\left( \frac{1}{2} +\frac{x_0}{L}\right) -
\psi^{\prime}\left( \frac{1}{2} -\frac{x_0}{L}\right)\right], 
\end{align}    
\noindent where a summation formula from Ref.~\onlinecite{prudnikov86integrals} was used
and $\psi(x)$ denotes the digamma function~\cite{Abramowitz}.

The energy of the point charge interaction with electrodes,  $U_q$, can be obtained by 
integration of the force (\ref{force-2el}) considering the relation 
$f(x_0)=-U_q^{\prime}(x_0)$. The integration constant $U_q(0)$ should be chosen so
that, in the limit $|x_0+L/2|\ll L$, the energy $U_q(x_0)$ reduces to the case
of a single interface, Eq.~(\ref{single-barrier}). Thus, the expression 
\begin{equation}
\label{Uq-form}
U_q(x_0) = \frac{q^2}{16 \pi \epsilon_0 \epsilon_s L}
\left[ \psi\left( \frac{1}{2} +\frac{x_0}{L}\right) +
\psi\left( \frac{1}{2} -\frac{x_0}{L}\right) + 2\gamma \right]
\noindent 
\end{equation}
\noindent results, where $\gamma=0.5772$ is Euler's constant~\cite{Abramowitz}.

\bibliographystyle{plain}
\bibliography{apssamp}

\end{document}